\documentclass[twocolumn,superscriptaddress,floatfix]{revtex4-1}
\usepackage{graphicx,epstopdf,amsmath}
\begin{document}

\newcommand{\ie}{\emph{i.e.}}
\newcommand{\eg}{\emph{e.g.}}
%
%
%
%
\newcommand{\FigCAL}{%
\begin{figure}[htbp]
\includegraphics*[width=\columnwidth]{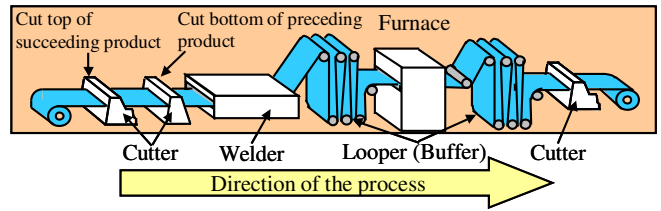}
\caption{(Color Online) The Continuous Annealing Line which is a core process for manufacturing diverse steel sheets.}
\label{fig_CALProcess}
\end{figure}}
\newcommand{\FigLongtermPlan}{%
\begin{figure}[htbp]
\includegraphics*[width=\columnwidth, height=55mm]{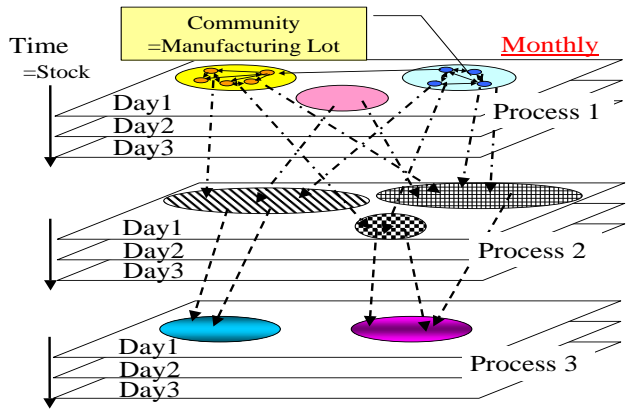}
\caption{(Color Online) Schematic illustration of long term multi-process planning.}
\label{fig_planning}
\end{figure}}
\newcommand{\FigOutlineofthisapproach}{%
\begin{figure}[htbp]
\includegraphics*[width=\columnwidth]{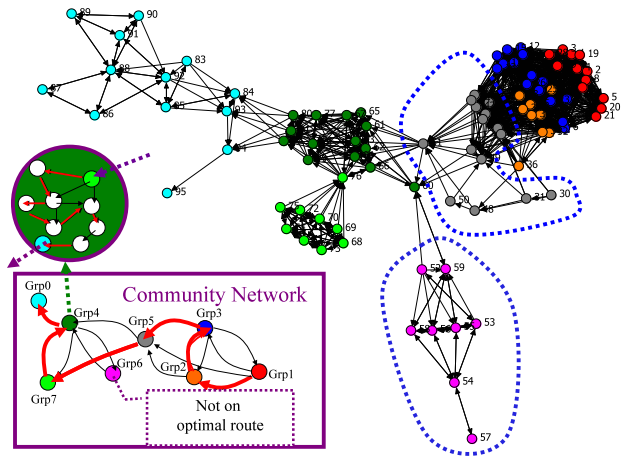}
\caption{(Color Online) Outline of proposed approach:  Nodes of the same color belong to a community~\cite{Leicht}.  Two {\em representative} communities encircled by blue loops are shown. 
Inset (bottom left) shows the extracted ``community network". Each node in this network (Group $0,1,2, ...,7$) represents one of the eight communities in the parent network and is denoted by a different color. Using $\Sigma$\{edge density along the route\}  as a performance metric, the solid red line on the community network shows the optimal route. As seen, this path does not pass through every community node, such as the one in magenta. The bigger dark green vertex shows the constituent nodes in the parent network explicitly, and the light green and cyan vertices are the start and end vertices respectively for the path through the community. A vertex which has high betweenness~\cite{Freeman} is expected to play a  key role here. }
\label{fig_outline}
\end{figure}}
\newcommand{\FigDegree}{%
\begin{figure}[htbp]
\includegraphics*[width=\columnwidth, height=55mm]{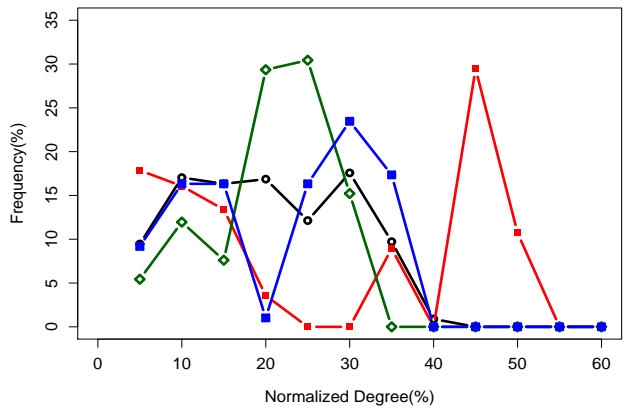}
\caption{(Color Online) Degree distributions of networks related to CAL. The red, green, blue and black  denote day $4$, day $6$, day $9$ and aggregate over all nine days respectively.}
\label{FigDegree}
\end{figure}}
\newcommand{\FigSolution}{%
\begin{figure}[htbp]
\includegraphics*[width=\columnwidth]{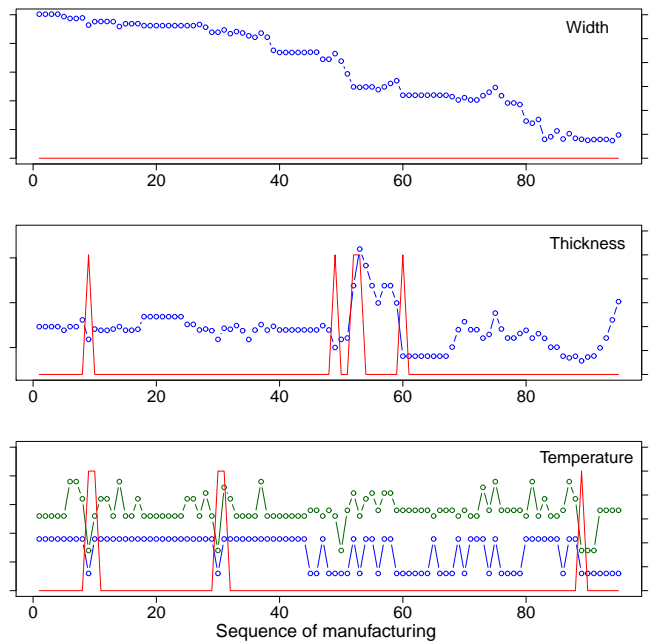}
\caption{(Color Online) A product schedule generated by our algorithm, with dots indicating the attribute value for successive products.  The y-axis on the left measures : width (blue) for top figure, thickness (blue) for middle figure, and, lower limits (blue) and upper limits (green) of annealing temperature for bottom figure (y-axis values not specified  for proprietary reasons). Spikes in the red line indicate instances of constraint violations.}
\label{FigSolution}
\end{figure}}
\newcommand{\FigComparison}{%
\begin{figure}[htbp]
\includegraphics*[width=\columnwidth, height=60mm]{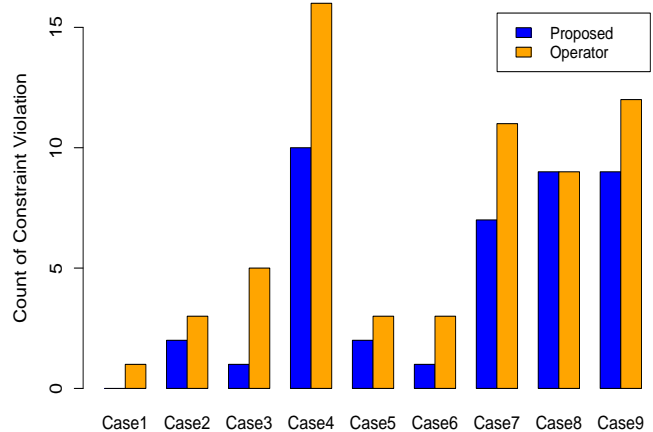}
\caption{ (Color Online) The comparison with operator's scheduling results obtained from 
real manufacturing process. Constraint violations which are highly undesirable 
are lessened by our proposed approach.}
\label{FigComparison}
\end{figure}}
\newcommand{\FigCorrelation}{%
\begin{figure}[htbp]
\includegraphics*[width=70mm]{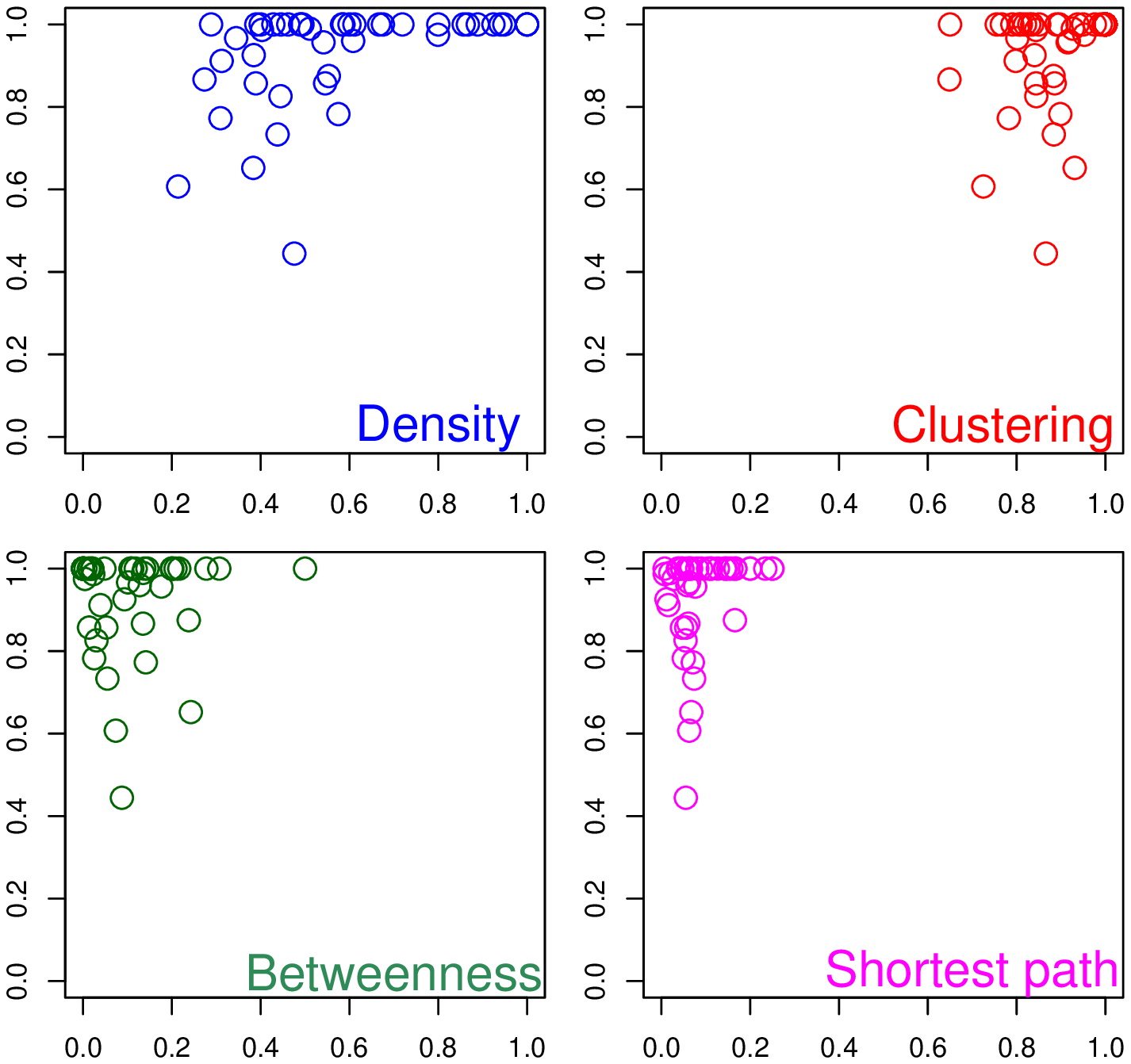}
\caption{(Color Online) Scatter plot between network metrics and difficulty in 
{\em long-term} scheduling (fraction of nodes in the longest Hamiltonian path).}
\label{FigCorrelation}
\end{figure}}
\title{Efficient scheduling using complex networks}
\author{Osamu Yamaguchi}
\email{o-yamaguchi@jfe-steel.co.jp}
\affiliation{JFE Steel Corporation, 1-1 Minamiwatarida-cho, Kawasaki-ku, Kawasaki,  210-0855, Japan}
\author{Soumen Roy}
\email{soumen@boseinst.ernet.in}
\affiliation{Bose Institute, 93/1 Acharya Prafulla Chandra Roy Road, Kolkata 700 009, India} 
\author{Raissa M. D'Souza}
\email{raissa@cse.ucdavis.edu}
\affiliation{University of California, Davis, CA 95616, USA}
\affiliation{Santa Fe Institute, 1399 Hyde Park Road, Santa Fe, New Mexico 87501, USA}

\begin{abstract}
We consider the problem of efficiently scheduling the production of goods for a model steel manufacturing company.  We propose a new approach for solving this classic problem, using  
techniques from the statistical physics of complex networks in conjunction with depth-first search to generate a successful, flexible, schedule.  The schedule generated by our algorithm is more efficient and outperforms schedules selected at random from those observed in real steel manufacturing processes.  Finally, we explore whether the proposed approach could be beneficial for long term planning. 
\end{abstract}

\maketitle
Operations research theories typically concentrate on deriving an optimal solution under fixed assumptions.  Such assumptions of fixed, constant conditions and constraints are often necessary to adequately simplify complex real life situations in theoretical treatment of problems. Yet, this constancy of conditions severely limits the applicability of simplified algorithms to important scientific and industrial issues, such as automated, dynamic, scheduling in various manufacturing processes across many industries. It is also notable that even with assumptions of invariable conditions, the computational complexity involved in combinatorial problems such as scheduling remains significant. To solve such combinatorial problems,  approximation methods like genetic algorithms or simulated annealing are often used~\cite{Eglese1990,GA1991}. 

In real-world manufacturing houses there is a continual and often unpredictable change of circumstances, ranging from calamities, to variable market demand, to labor strikes. A prime example of a complex constraint influencing scheduling is the storage time of finished products which is in turn dictated by  variable market demand. The cost of prolonged storage of completed products is prohibitive for industries who also  need to be flexible to accommodate rush production requests which can command premium pricing.

This letter proposes a new approach for scheduling and planning of various manufacturing processes using  techniques based on the statistical physics of complex networks~\cite{BarabasiAlbert,NewmanReview,Barrat2008,NewmanBook2010, epl091} 
and classical depth-first search.   We conceptualize a model steel plant by groups of products forming networks where  each distinct product is represented by a vertex and 
{\em directed edges} connect products than can be manufactured in succession as dictated by manufacturing constraints. We call this the product network.  Our proposed approach draws connections between statistical physics of complex networks and scheduling of manufacturing  processes and generates significant advantages.  Firstly, it allows a rapid calculation of an efficient daily schedule as shown herein. Secondly,  the network approach could in principle lead to {\em approximate} scheduling solutions for relatively longer times like weeks or months where more precise traditional solutions can not be computed due to the complexity of the algorithms and increased size of the space of products that could be produced in this longer time frame.  
Thirdly, for strategic decisions, like investments to increase productivity, it is crucial to identify the improvements to be gained by reforming the constraint network, modifying it in a way which is most efficient from the manufacturing point of view. Fourthly, at an operational level,   experience has shown that visual information of the network  structure of products and constraints is immensely helpful to human operators, especially less experienced operators, overseeing the manufacturing process. Lastly, a clear knowledge of the network is of immense help when quick and decisive intervention to the schedule  is required in real time or in an emergency.

\FigCAL%
\FigLongtermPlan%
 In many steel manufacturing processes, efficiency in productivity is achieved by welding subsequent products together.  For example, a continuous annealing line (CAL) shown in Fig. ~\ref{fig_CALProcess}, allows for production of a continual stream of products from cold rolled steel sheets while achieving a desired malleability and ductility. For welding and a continuous temperature transition between products, three attributes are very important in CAL process: {\em the width of the steel sheet,  its thickness and the annealing temperature}. The first two are constraints for welding and determine the directionality of an edge connecting together two products in the product network. For width, a roughly decreasing transition is suitable; while for thickness, only a permitted range of thickness changes between two products is allowed. The third constraint imposes the condition that any two given products produced in sequence 
 should have a common range of annealing temperatures allowing for a smooth change of temperature settings in the furnace.

During production, the steel products attain their final shape as a consequence of various manufacturing processes such as casting, hot rolling, cold rolling, galvanizing,  etc.  When groups of products are subject to similar manufacturing conditions they are classified together into a {\emph {manufacturing lot}. Forming a manufacturing lot can be quite complicated as reflected in Fig. \ref{fig_planning}. If a daily or process specific grouping is too small to be  considered a manufacturing lot, it is possible to extend the time-axis of Fig.~\ref{fig_planning} by a small amount to thus increase lot size and productivity.   Since multiple manufacturing processes are involved over longer time frames, a traditional operations research approach  becomes too complex to implement.  Therefore, applying network  theory methods~\cite{BarabasiAlbert,NewmanReview,Barrat2008, NewmanBook2010,epl091} is an attractive alternative approach.

\FigOutlineofthisapproach
The treatment of a diverse range of complex real-world systems as networks has resulted in many advances in the past decade~\cite{BarabasiAlbert,NewmanReview,Barrat2008,NewmanBook2010, epl091}.  One particularly active area of research involves the {\em community structure} of a network, or how a network breaks up into sub-groups of nodes with dense connectivity within a group, but only sparse connections between groups~\cite{GirvanNewman}.  Such community detection algorithms, including those for large networks, concepts of overlapping versus non-overlappinng communities, and many additional features, are extensively covered in a recent comprehensive review~\cite{Fortunato2010}.  For our purposes, we expect that the community detection algorithm for {\em directed networks} will downsize the magnitude of the scheduling problem for both daily scheduling and complex long-term planning. We eventually want to find the relation between the properties of networks as quantified by standard metrics~\cite{epl091} and the difficulty of scheduling, with preliminary results included at the end of this letter. For instance, if the product network has low clustering \cite{Watts-Strogatz} ({\it i.e.,} small numbers of transitive triangles) and vertices of low degree ({\it i.e.,} low connectivity), we expect that scheduling will be difficult.

First, we attempt a solution to daily scheduling which satisfies all manufacturing constraints. 
We define the difficulty of scheduling as the probability of existence of a {\em Hamiltonian path} in the network. 
A Hamiltonian path on a network is a route which passes through each vertex in the  network {\em exactly once}. To construct a Hamiltonian path (which is also an efficient path as shown below), we propose a new algorithm for CAL scheduling, based on the community detection algorithm for a directed network discussed in Ref.~\cite{Leicht}.  Fig. \ref{fig_outline} illustrates the algorithm  which is outlined in the following steps: 
\noindent
(1) Network construction: Build a basic network where a vertex represents a product and an edge denotes two products can be connected together by welding in CAL process. 
\noindent
(2) Add weights to the edges according to the width change between products:  each weight is set to one of $\{$99, 50, 2, 1$\}$  with higher weights assigned to more desirable transitions.   It must be noted that  transitions from wide to narrow 
 are strongly preferred in CAL finishing lines.  We categorize each product by its width in $100{\rm mm}$ pitch. The weight $99$ is given to an edge when both vertices belong to the same width category. The weights $50$ and $2$ are assigned for transitions from a wider to a narrower product that are one and two categories distant from each other respectively. The smallest weight is assigned to edges which have a narrow to wide width transition. 
\noindent
(3) Community detection: 
We identify community sub-graphs and build a coarser-grained {\emph {community network}}, where each vertex represents a community and a edge means a connection between communities. Every edge has a weight which represents the connection density between two communities.  
\noindent
(4) Find an optimal route which  includes the maximum number of communities: We use $\Sigma$\{edge density along the route\}  as a performance index and the solid red line on the community network in Fig.  \ref{fig_outline},  shows an optimal route passing through the communities. 
\noindent
(5) Select a start vertex and an end vertex in each community: In Fig.  \ref{fig_outline}, the light green vertex and cyan vertex in the big dark green circle denote start and end vertices, respectively. Note that a vertex which has high betweenness  centrality is expected to be a key vertex for the solution. Betweenness centrality of a node indicates the relative fraction of shortest paths between all nodes that pass through this given node~\cite{Freeman}.  
\noindent
(6) Find a route passing inside of community from the specified start vertex to the end vertex: We apply depth first search to obtain a Hamiltonian path. Obviously, the traveling salesman problem could be applied if we need to optimize some performance index. 
\noindent
(7) We can construct the solution from (4) and (6) and if there are any vertices which are not included in combined solution, we use the cheapest insertion method and obtain a complete solution. 
\FigDegree%
\FigSolution%

Next, we validate our methods using nine days of daily production data, obtained from a  CAL at JFE steel~\cite{JFE}. Our networks vary from  having $70$ to $112$ vertices and from $675$ to $1535$ edges. Fig. \ref{FigDegree} shows the degree distribution for three different {\em representative} days of this data and for the aggregate data over all nine days. The  difference between the distributions of the individual days  implies that the networks at hand seem to be non-self-averaging~\cite{pla061} and it is not easy to find a simple or general solution.  

\FigComparison%
A scheduling solution obtained by the algorithm proposed above is shown in Fig. \ref{FigSolution}, plotting the attribute values of products manufactured in succession via this schedule. The top, middle and bottom plots represent the width, thickness and annealing temperature respectively, and hence the required transitions between subsequent products.  Spikes in each figure represent constraint violations.  If there exists a constraint violation,  in practice on the CAL, this means a section of steel sheet is introduced which is not machined into a product, but included only to implement the necessary physical constraints.  Analogously, in our product network, we introduce a dummy node which is not an actual product but is inserted for the explicit purpose of satisfying the constraints. Thus, each constraint violation directly reduces productivity and is obviously highly undesirable. Moreover, for any algorithm to succeed, the number of constraint violations it introduces, must be minimal.
 
\FigCorrelation%
Figure \ref{FigComparison} compares the number of constraint violations observed in practice to the number that result from our scheduling algorithm.  Clearly the solutions obtained by our proposed algorithm are generally better than the results obtained  manually by the operator on the workshop  floor. This implies that the community detection algorithm is quite useful for downsizing the original scheduling problem and deriving a useful solution.

Inspired by the success of our approach for daily production results,  we apply our community detection algorithm to nine days of production data, where each day's data divides into $3$ to $6$ communities, with the total adding upto $45$ communities. More generally we are interested in identifying whether common network metrics like edge density, clustering co-efficient, betweenness centrality, and average shortest path can be related with the difficulty of {\em long-term scheduling}.  We quantify the latter, akin to our previous definition, by the fraction of  nodes included in the longest Hamiltonian Path of the network.  As shown in the Fig. \ref{FigCorrelation}, there is no simple linear relation between the four major  network metrics and  the  fractional size of Hamiltonian Path for these $45$ communities. In fact, $R^2$ for a linear regression between these four properties of communities and Hamiltonian path is $0.14$.

It is well-known that the Hamiltonian Path problem is NP-complete and therefore it is non-trivial to find an approach which would simplify the problem. We hope that future research will  lead to simple relations between network metrics or properties (which are undiscovered as yet), and the difficulty of scheduling for a long term, large scale planning problem.

In summary,  we aim to find a realistic solution to the problem of automated scheduling which takes into account changing constraints observed in actual production processes. We propose an algorithm using  community detection methods based on the statistical physics of networks. The approach proposed herein successfully derives a solution by downsizing the original problem. For daily scheduling, the results obtained our algorithm are  better than those practiced on the workshop floor. We show that it is indeed difficult to derive long-term schedules on account of the computational complexity involved. We hope that the approaches proposed in this work will lead to the creation of new network metrics or algorithms which will effectively address the challenges of long-term scheduling.

\end{document}